\documentclass[10pt]{iopart}

\usepackage{hyperref}
\usepackage{graphics}
\usepackage{epsfig}
\usepackage{subfigure}
\usepackage{color}

\begin{document}

\title{Transitions in a Probabilistic Interface Growth Model}

\author{S G Alves$^1$ and J G Moreira$^2$}
\address{$^1$ Departamento de F\'{\i}sica, Universidade Federal de
Vi\c cosa, 36570-000, Vi\c cosa, MG, Brazil.}
\ead{sidiney@ufv.br}
\address{$^2$ Departamento de F\'{\i}sica, Universidade Federal de
Minas Gerais, Caixa Postal 702, 30161-970, Belo Horizonte, MG, Brazil.}
\ead{jmoreira@fisica.ufmg.br}

\begin{abstract}
We study a generalization of the Wolf-Villain (WV) interface growth model based on a probabilistic growth rule. In the WV model, particles are randomly deposited onto a substrate and subsequently move to a position nearby where the binding is strongest. We introduce a growth probability which is proportional to a power of the number $n_i$ of bindings of the site $i$: $p_i\propto n_i^\nu$. Through extensively simulations, in $(1+1)$-dimensions, we find three behavior depending of the $\nu$ value:
{\it i}) if $\nu$ is small, a crossover from the Mullins-Hering to the Edwards-Wilkinson (EW) universality class;
{\it ii}) for intermediate values of $\nu$, a crossover from the EW to the Kardar-Parisi-Zhang (KPZ) universality class;
{\it iii}) and, finally, for large $\nu$ values, the system is always in the KPZ class.
In $(2+1)$-dimensions, we obtain three different behaviors:
{\it i}) a crossover from the Villain-Lai-Das Sarma to the EW universality class, for small $\nu$ values;
{\it ii}) the EW class is always present, for intermediate $\nu$ values;
{\it iii}) a deviation from the EW class is observed, for large $\nu$ values.
\end{abstract}

\journal{ JSTAT}
\maketitle

\section{Introduction}

As the interface growth is ubiquitous in nature, it have constituted an important subject of research in the past decades~\cite{Barabasi,Meakin}. In general, the thickness of these interfaces, also named roughness $w$, is a signature of the nonequilibrium growth conditions under which the interface has formed and their evolution can be characterized by the Family-Vicsek scaling relation \cite{fv}
\begin{equation}
w(L,t)=L^{\alpha}f\left({t\over L^z}\right)~,
\end{equation}
where $L$ is the linear size of the system and $t$ is the evolution time. The scaling function $f(x)$ behaves as
\begin{equation}
f(x) \sim
\left\{ \begin{array}{ll}
x^\beta, & \mbox{for } x\ll 1 \\
\mbox{const}, & \mbox{for } x\gg 1 
\end{array}\right.
\end{equation}
where $\beta =\alpha/z$. Then, the roughness $w$ grows with $t^\beta$ until saturates in a value $w_{sat}$ denominated saturation roughness, which behaves with the system size $L$ as a power law according as $L^\alpha$. The saturation times $t_{sat}$ also grows with the system size as $L^z$. The exponents $\alpha,~\beta$ and $z$ are known as roughness, growth and dynamic exponents, respectively.

This scaling analysis lead to a considerable advance in the understanding of the roughening of interface growth and allow to define some universality classes which are related to a stochastic differential equation for the height $h(\vec x,t)$ of the system at position $\vec x$ and time $t$. Lai and Das Sarma \cite{Lai} propose a general form for these equations
\begin{equation}
  {\partial h\over\partial t}= \nu_0\nabla^2 h + \nu_1\nabla^4 h + \ldots +
  \lambda_0(\nabla h)^2 + \lambda_1(\nabla^2 h)^2 +
  \lambda_2 \nabla\cdot(\nabla h)^3 +\ldots +\eta~,\label{eq_gen}
\end{equation}
where the parameters $\nu_i$ are related to the linear terms, $\lambda_i$ to the nonlinear ones and $\eta$ is a white noise.

In the hydrodynamic limit ($L\rightarrow\infty$ and $t\rightarrow\infty$), the $\lambda_0$ term dominates and the asymptotic behavior is governed by the KPZ universality class \cite{kpz}. This equation is the simplest nonlinear differential equation to describe a kinetic growth process and it has exact solution only in $(1+1)$-dimensions where the scaling exponents are $\alpha =1/3$ and $z=3/2$.

For models where the up-down symmetry is present, $\lambda_0 =0$ \cite{Barabasi} and the $\nu_0$ term of equation (\ref{eq_gen}) dominates. In the asymptotic limit, we recuperate the linear equation proposed by Edwards and Wilkinson \cite{ew} to studied a sedimentation process. This equation has an exact solution and the scaling exponents are given by $\alpha =(2-d)/2$, where $d$ is the dimension of the substrate, and $z=2$.

However, for short and intermediate times, the process can present a behavior which is a portrait of the dominant term before of the hydrodynamic limit. An example is the linear equation proposed by Wolf and Villain \cite{wv} and by Das Sarma and Tamborenea \cite{sarma-tamborenea} to introduce surface diffusion where the term $\nu_1$ dominates for short times. Equation (\ref{eq_gen}) just with these term is linear and the exact solution gives $\alpha =(4-d)/2$ and $z=4$.
In table 1 we show the growth and the roughness exponents associated to the main terms of equation (\ref{eq_gen}), in $d=1$ and $d=2$.

\begin{table}[!h]
\caption{The growth $\beta$ and the roughness $\alpha$ exponents associated to the main terms of equation (2) (left column) for $d=1$ (center) and $d=2$ (right)~\cite{Barabasi}.}
\begin{center}
\begin{tabular}{|c|cc|cc|} \hline
                          & $\beta$  & $\alpha$   & $\beta$     & $\alpha$ \\ \hline
 $\nabla^2 h$             & $1/4$    & $1/2$      & $0$ ($log$) & $0$ ($log$) \\
 $(\nabla h)^2$           & $1/3$    & $1/2$      & $\approx 0.24$ & $\approx 0.4$ \\
 $\nabla^4 h$             & $3/8$    & $3/2$      & $1/4$       & $1$         \\
 $\nabla^2 (\nabla h)^2 $ & $1/3$    & $1$        & $1/5$       & $2/3$      \\
 $\nabla \cdot (\nabla h)^3 $   & $3/10$    & $3/4$    & $1/6$  & $1/2$   \\ \hline
\end{tabular}
\end{center}
\end{table}

The determination of the asymptotic universality class of a growth model may be a difficult task due to the extremely long time of the initial behavior that can mask the crossover to another one \cite{tales}. To investigate this problem various techniques have been developed, as examples we can cite the investigation of the universality class using the surface diffusion currents \cite{Krug, Krug1}, the application of the noise reduction technique to models with limited mobility \cite{Das_NR}, the investigation of finite size effects in the DT and WV models \cite{Costa}, the application of the scaling  transformation to the stochastic equation of the WV model \cite{Vvedensky,Vvedensky1}.

In this article, we propose a probabilistic interface growth model that is a generalization of the WV model where an adatom diffuses to maximize its coordination number.
In our model, the incoming particle search for a site where the number of bonds is higher and it has a probability to be incorporate in this site. The probability depends of the number of bonds that the particle will have and it has a parameter $\nu$ so that, for $\nu=0$, the WV model is recuperated. For low values of $\nu$, the crossover to the asymptotic universality class occurs early and, for high values, the non-linear KPZ universality class appears. A similar strategy was usaed in the investigation of the on-lattice Eden model \cite{Paiva}, where the authors show that the introduction of such probability growth leads to the growth of on-lattice Eden clusters whithout the undesirable anisotropy effects. This strategy was initially proposed to generate an isotropic cluster of the on-lattice diffusion limited aggregation model \cite{Bogoyavlenskiy}. However, a later work \cite{Alves_bogo} shows that in the noiseless limit, instead of isotropic patterns, a $45^\circ$ ($30^\circ$) rotation in the anisotropy directions of the clusters grown on squre (triangular) lattices was observed.
The model that we study is presented in the next section and the results in section 3. The last section shows some conclusions.
\section{Probabilistic Growth Rule}

We study a model where the particles are randomly deposited on a $d$-dimensio\-nal lattice of linear size $L$. The height of the interface at time $t$ is represented by  $h_i(t)$ and the number of bonds of each site by  $n_i(t)$, with $i = 1, 2, \ldots, L^d$.  The initial condition is given by $h_i(0) = 0$ and $n_i(0) = 1$, $\forall i$, {\it i.e.}, at the beginning of the simulation the interface is flat and a particle deposited into the site $i$ will have $n_i=1$ bonds, $\forall i$. A particle is deposited at random and search in its neighborhood for the site with the largest number of bonds and this site is chosen as the growing site. If the number of bonds of the particle cannot be increased in the neighborhood, the particle will choose the deposition site as the growing site and, if it has more than one site with the same number of bonds, one of them is chosen at random. One particle is incorporated to the growing site with a probability given by
\begin{equation}
p_i=\left(\frac{n_i}{n_{max}}\right)^\nu
\end{equation}
where $n_{max} = 2 d + 1$ is the maximum number of bonds of one site, that is the case of a particle deposited in a site with its neighborhood fully occupied in a $d$-dimensional lattice. The variable $\nu$ is the parameter of the model that control the growth probability of a given site and, for $\nu = 0$, the WV original growth model is recovered.
Note that the rate growth of the sites with smaller neighbors number are decreased while the growth rate of sites with $n=n_{max}$ is always equal to $1$ and an increase in the control parameter $\nu$ emphasize the difference of these probabilities.
After the deposition of a particle, the neighbors number of the growth site and of its neighbors must be updated. As usual \cite{KK}, we define one time unit as the tentative of deposition of $L$ particles.

\section{Results}

\begin{figure}[!b]
\begin{center}
\epsfxsize=14cm\epsfbox{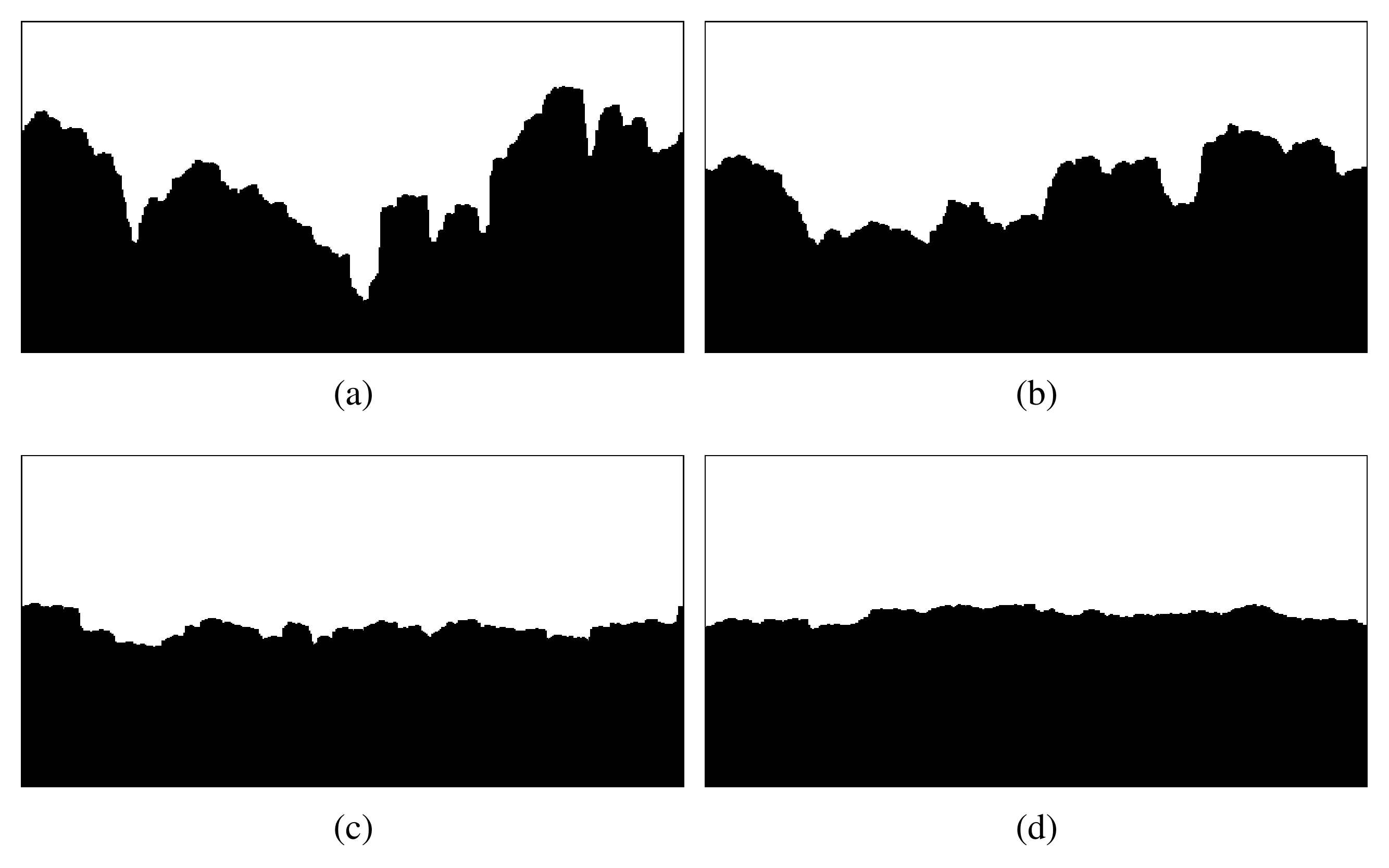}
\end{center}
\caption{\label{wv_perfis} \footnotesize Interface morphology obtained after $10^5$ times step in a $(1+1)$-dimensions lattice of size $L = 400$. We used in these simulations $\nu = 0.0$, $0.1$, $0.50$ and $2.0$ for (a), (b), (c) and (d), respectively.}
\end{figure}

\begin{figure}[!b]
\begin{center}
\subfigure[\label{p2d_nu00}]{\epsfxsize=6cm\epsfbox{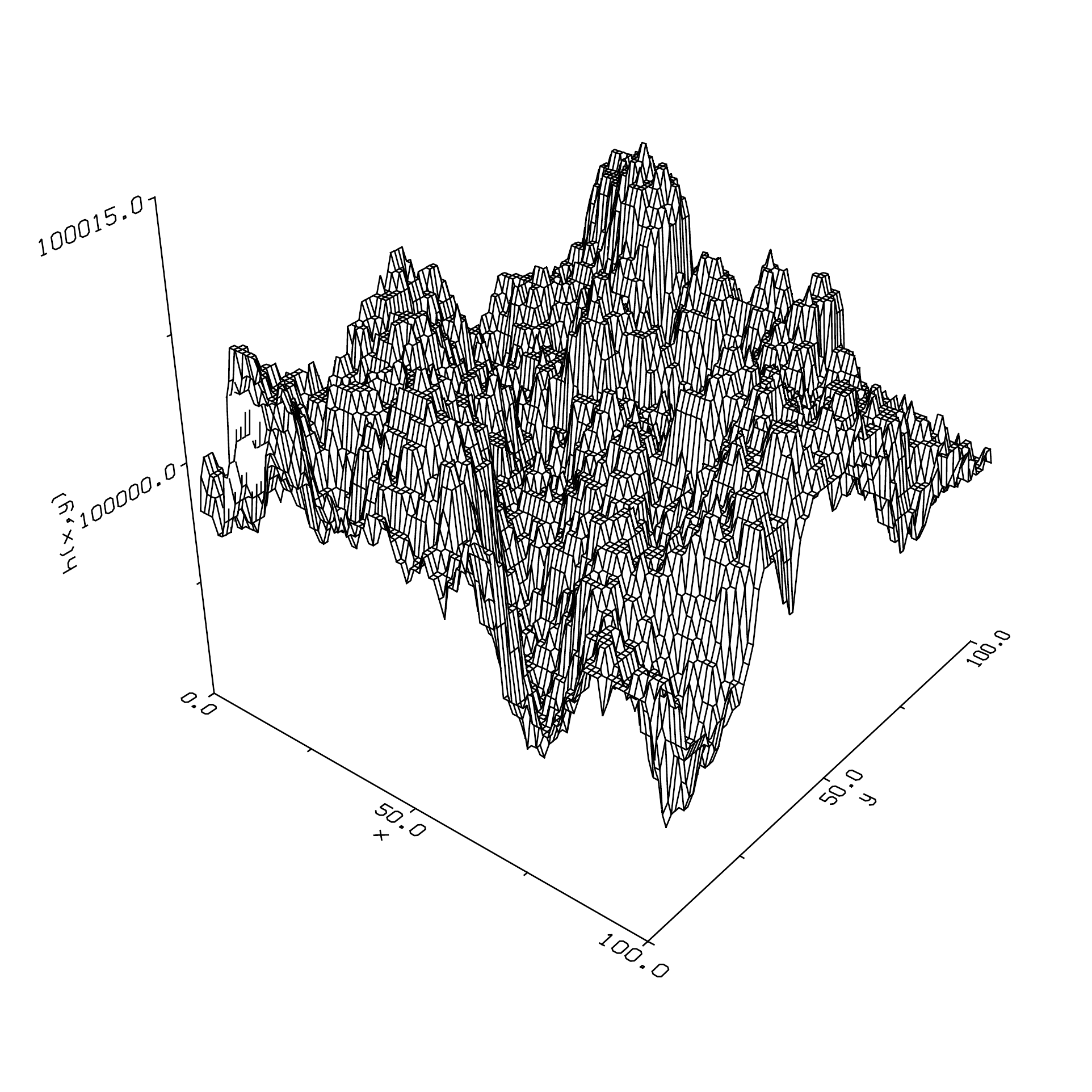}}
\subfigure[\label{p2d_nu20}]{\epsfxsize=6cm\epsfbox{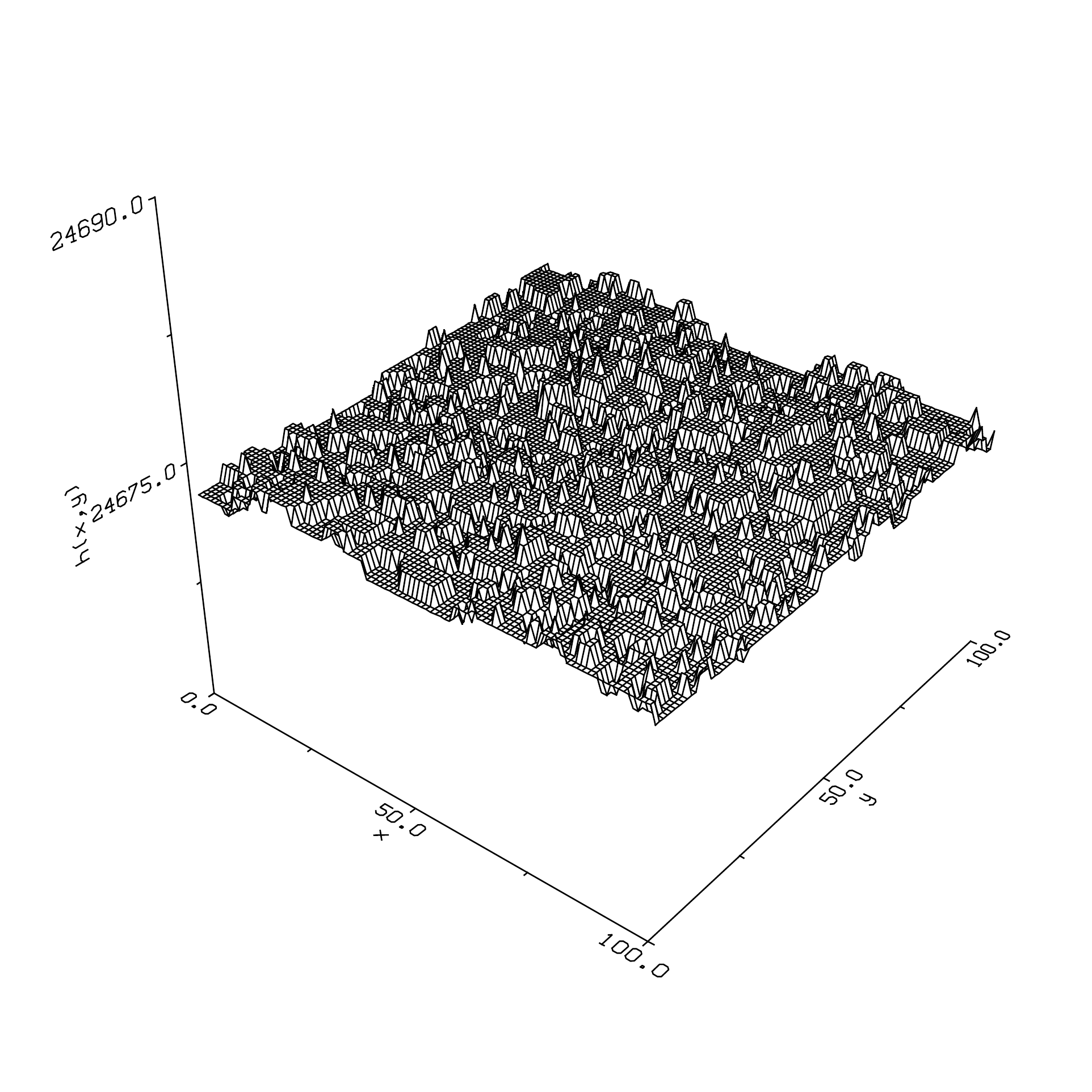}}
\subfigure[\label{p2d_nu40}]{\epsfxsize=6cm\epsfbox{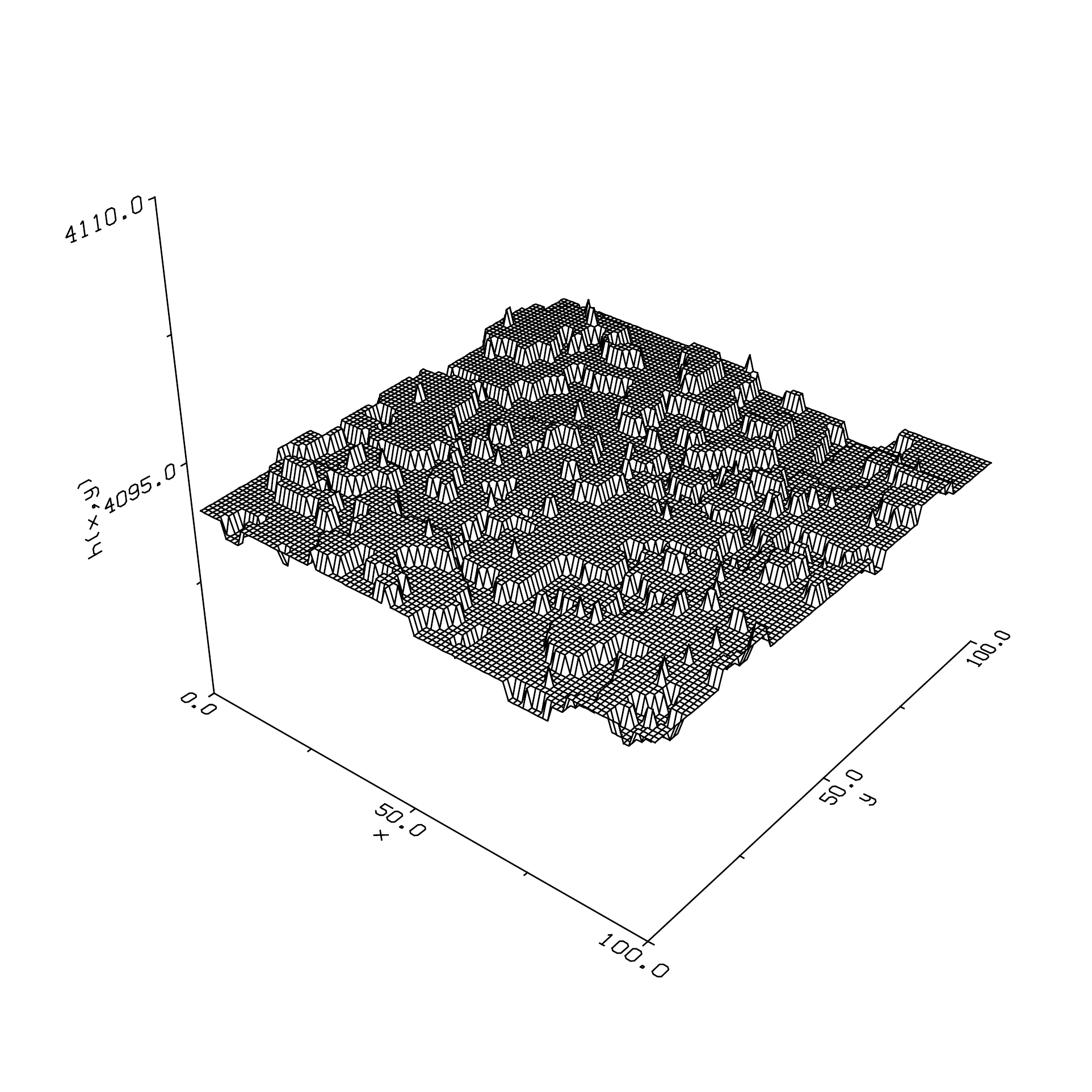}}
\end{center}
\caption{\label{p2d_nus}\footnotesize Interfaces obtained for distinct $\nu$ values in a $(2+1)$-dimensions system of linear size $L=100$. We used $\nu = 0.0$ in (a), $\nu = 2.0$ in (b) and $\nu = 4.0$ in (c).}
\end{figure}

As mentioned previously, if $\nu=0$, the WV model is recovered as we can see in figures~\ref{wv_perfis}(a), for $d=1$, and \ref{p2d_nus}(a), for $d=2$.
The interface obtained using this value exhibits a landscape governed by large plateaus separated by deep valleys that is the signature of models that incorporate diffusion~\cite{wv,sarma-tamborenea}. For small values of $\nu$, for example $\nu = 0.1$ as shown in figure~\ref{wv_perfis}(b), the interface resembles to that of ~\ref{wv_perfis}(a), however an attenuation of the largest steps are observed. Moreover, as can be seen from figure~\ref{wv_perfis}(c) and (d) and in figure~\ref{p2d_nus}(b) and (c), an additional increase in the $\nu$ value leads to an increase on this attenuation until the ridgeline landscape disappear completely, even for larger systems sizes and times asymptotically large.
Therefore, in both $(1+1)$ and $(2+1)$-dimensions, the effect of the $\nu$ parameter is to smooth the interface and this effect can be easily understood through an analyze of the growth probability. As observed previously, when we increase the $\nu$ parameter, the rate growth of sites with $n<n_{max}$ decreases, hence the appearing of peaks are reduced.

In order to characterize quantitatively the dependency of the roughening process by the parameter $\nu$, we consider the roughness of the interface that consist in the determination of the root mean square fluctuation in its height, {\it i.e.},
\begin{equation}
w(t) = \sqrt{\frac{1}{N}\sum_{i=1}^{N}\left(h_i(t)-\bar h(t)\right)^2}
\end{equation}
where $N=L^d$ is the number of sites in the substrate, $h_i(t)$ is the height of the site $i$ at time $t$ and $\bar h(t)$ is the mean height of the interface~\cite{Barabasi}. We carried out simulations in systems of size ranging from $L=10$ up to $10^5$ for $d=1$ while, for $d=2$, we used systems from $L=10$ up to $500$. The $\nu$ parameter was varied from $0$ to $8$ and from $0$ to $4$, for $(1+1)$ and $(2+1)$-dimensions, respectively.

\begin{figure}[!b]
\begin{center}
\subfigure[\label{rugo_L1e5A}]{\epsfxsize=6.5cm\epsfbox{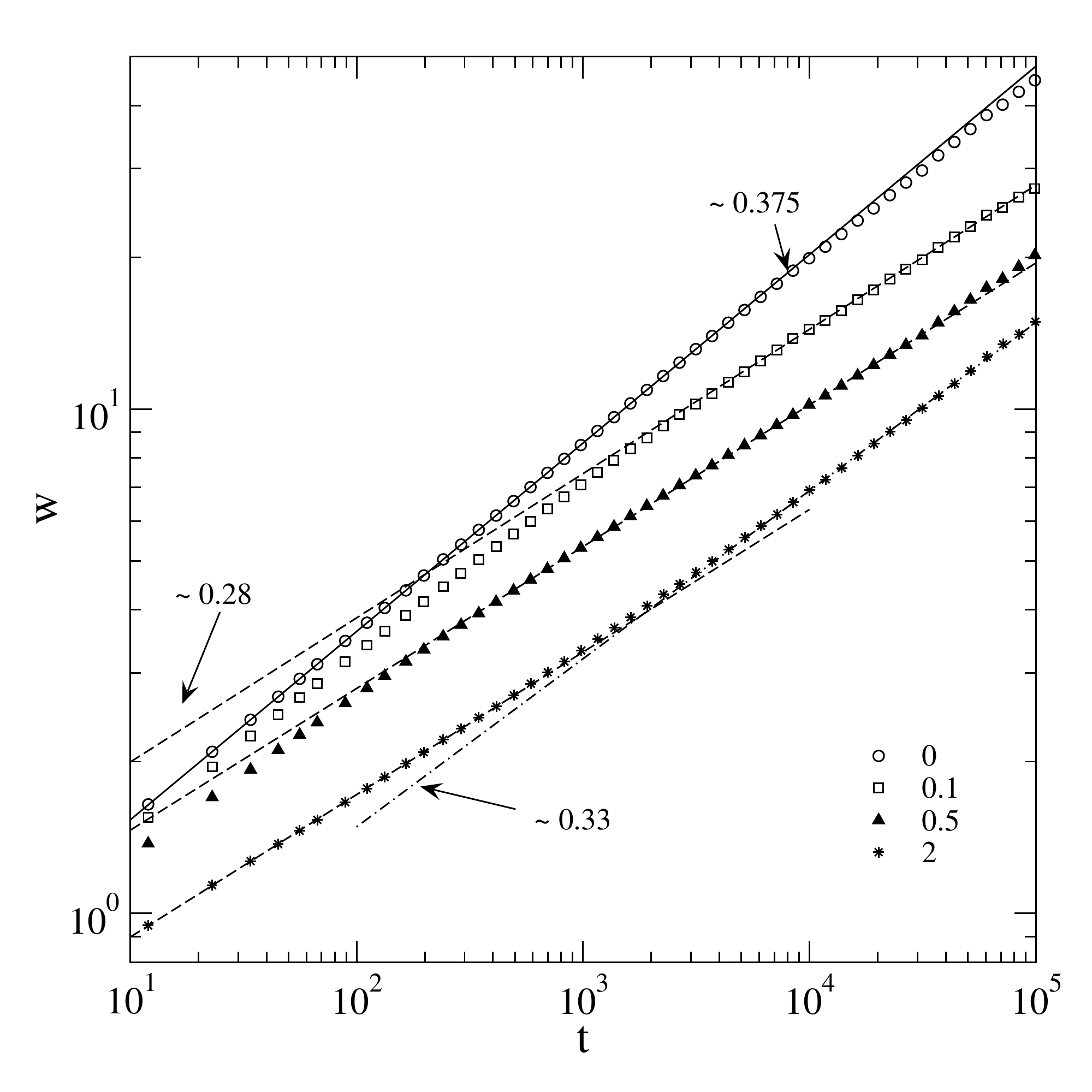}}
\subfigure[\label{rugo_L1e5B}]{\epsfxsize=6.5cm\epsfbox{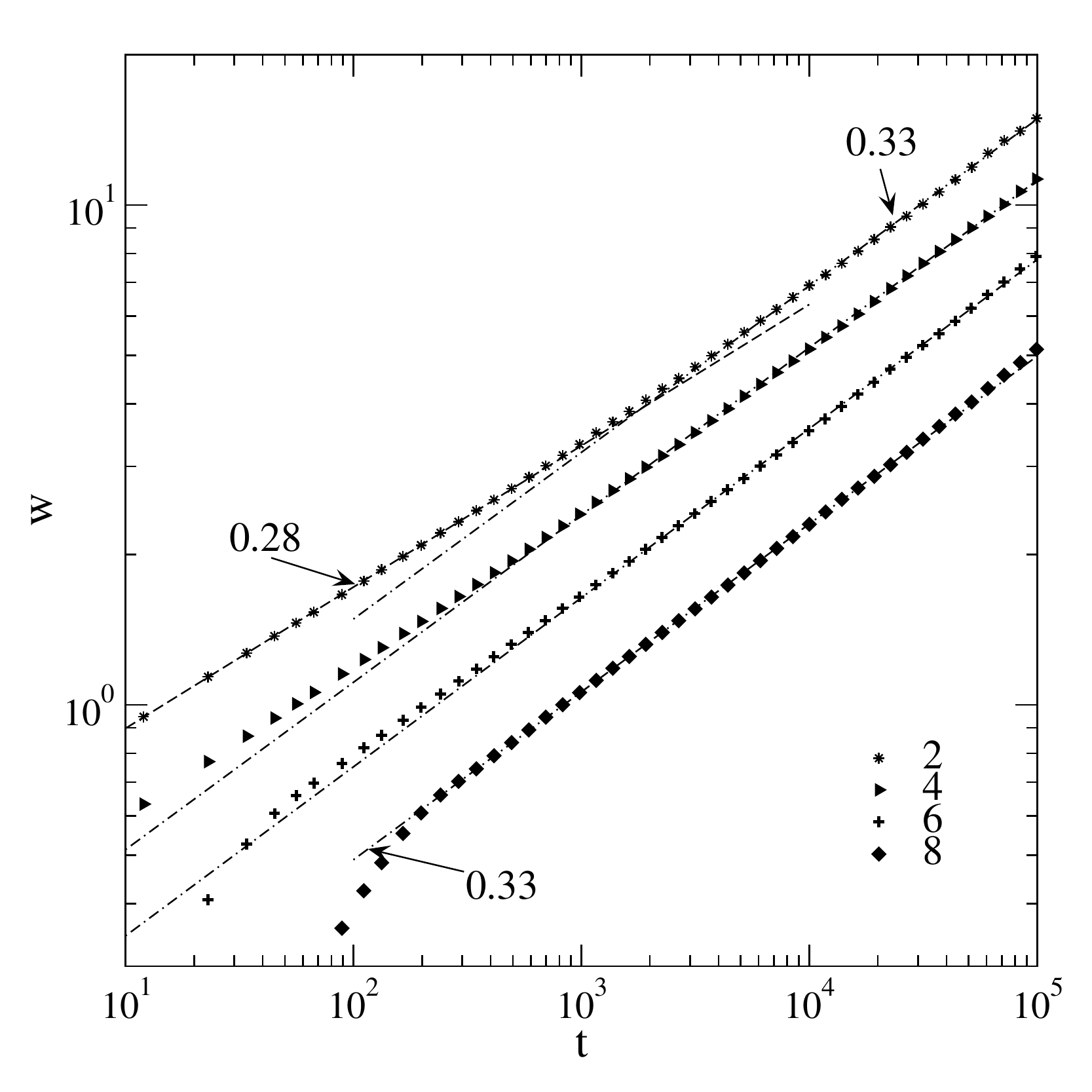}}
\end{center}
\caption{\label{ws} Temporal roughness evolution for a $(1+1)$-dimensions system of size $L=10^5$ for various values of the parameter $\nu$ as indicated in the legend. Notice that, to improve the visualization, the results for $\nu \le 2$ are show in (a), while for $\nu \ge 2$ are show in (b).}
\end{figure}

As we can observe in figure~\ref{ws}, for system in $(1+1)$-dimensions, an important consequence of the introduction of the growth probability is the arising of different behaviors.
If the $\nu$ value is set to zero (upper curve in figure~\ref{rugo_L1e5A}), we find $\beta \approx 0.375$ throughout all simulation, which means that the $\nabla^4h$ term of equation (\ref{eq_gen}) dominates.
However, for sufficiently small $\nu$ values ($\nu = 0.1$),  we observe a crossover in the $\beta$ exponent from a value $\approx 0.375$ to a value $\approx 0.28$, very close to $\beta = 1/4$ which means that the $\nabla ^2$ term of equation (\ref{eq_gen}) begins to prevail. This crossover is previously observed for the WV original model \cite{Krug,Ryu}.
For a small increase in the $\nu$ parameter, e.g. for $\nu = 0.5$ in figure~\ref{rugo_L1e5A}, the growth exponent is $\approx 0.28$ at all simulation.
However, for an further increase in the $\nu$ parameter ($\nu = 2.0$), we observe the exponent $\beta \approx 0.33$ after a crossover time.
Finally, for largest $\nu$ values (three bottom curve in figure~\ref{rugo_L1e5B}), we can observe $\beta \approx 0.33$ at all simulation. This new asymptotic growth exponent close to $1/3$ is directly related to the rejection of that particle which choose a site with a very low growth probability. Therefore, this asymptotic value can be associated to the KPZ universality class, similar to the restricted solid-on-solid model proposed by Kim and Kosterlitz \cite{KK}.

\begin{figure}[!h]
\begin{center}
\epsfxsize=6.25cm\epsfbox{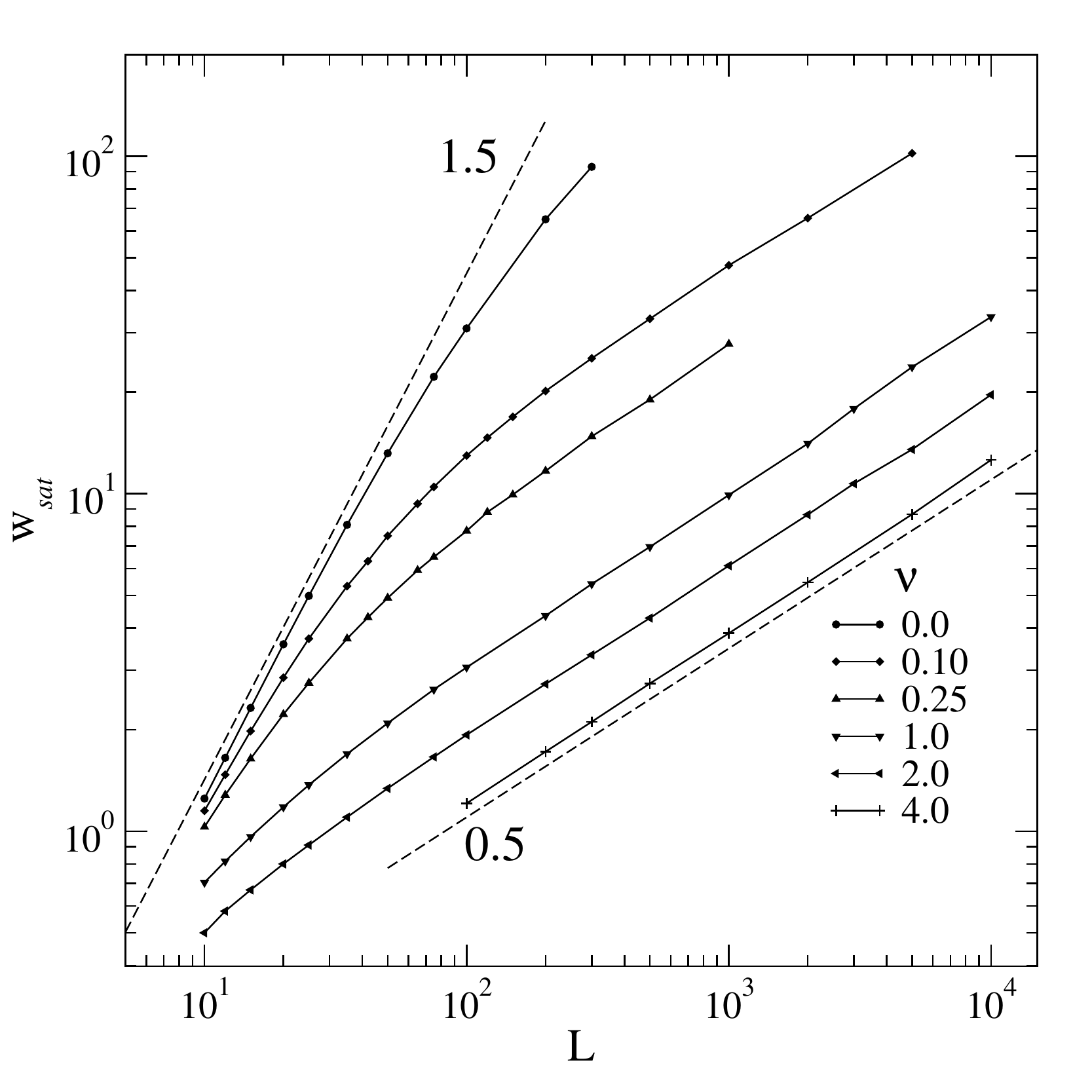}
\end{center}
\caption{\label{wsat_nus} Saturation roughness as function of the system size obtained for a $(1+1)$-dimensions system with distinct $\nu$ values indicated in the legend of the figure.}
\end{figure}

Corroborating the scaling of the roughness evolution, the saturation roughness also exhibits a crossover between different roughness exponents, as we can see in figure~\ref{wsat_nus}. When low values of the $\nu$ parameter are used, for small values of $L$, the roughness exponent exhibits a value close to $3/2$, as observed for the $\nabla^4h$ term, and a value close to $1/2$ for largest values of the system size $L$. However, when large $\nu$ values are used, this crossover disappear and the observed roughness exponent is always close to $1/2$, a value obtained for both the EW ($\nabla^2h$ term) and KPZ ($(\nabla h)^2$ term) universality classes.

\begin{figure}
\begin{center}
\subfigure[\label{w2d_nus}]{\epsfxsize=6.5cm\epsfbox{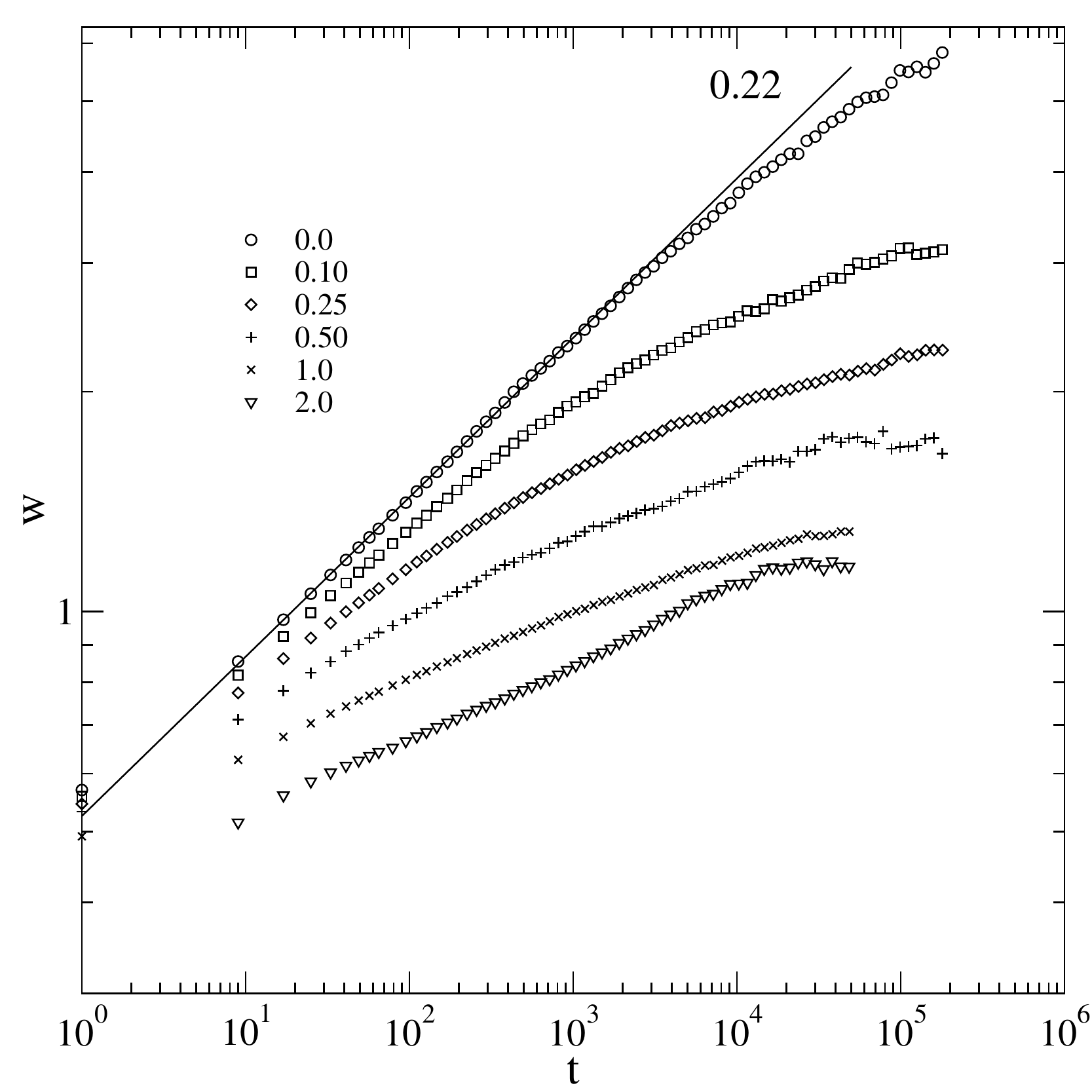}}~~~~~~~
\subfigure[\label{w2d_nuslog}]{\epsfxsize=6.5cm\epsfbox{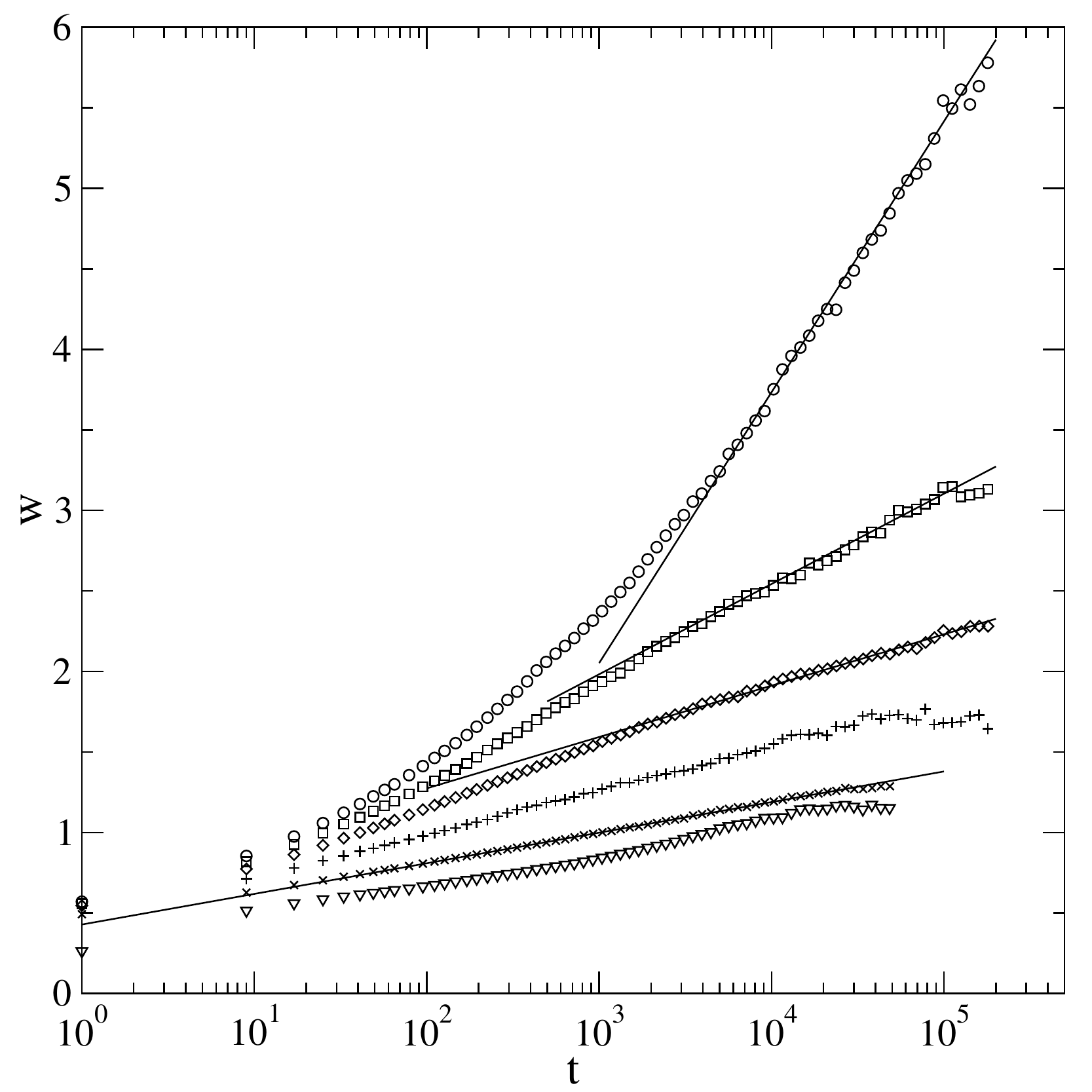}}
\end{center}
\caption{\label{rugo2d}\footnotesize Roughness time evolution for distinct $\nu$ values ($0.0$, $0.1$, $0.5$, $1.0$ e $2.0$) for a $(2+1)$-dimensions system of linear size $L=500$. In (a), a log-log plot is shown and, in (b), a semi-log plot.}
\end{figure}

\begin{figure}
\begin{center}
\subfigure[\label{wsat2d_nus}]{\epsfxsize=6.5cm\epsfbox{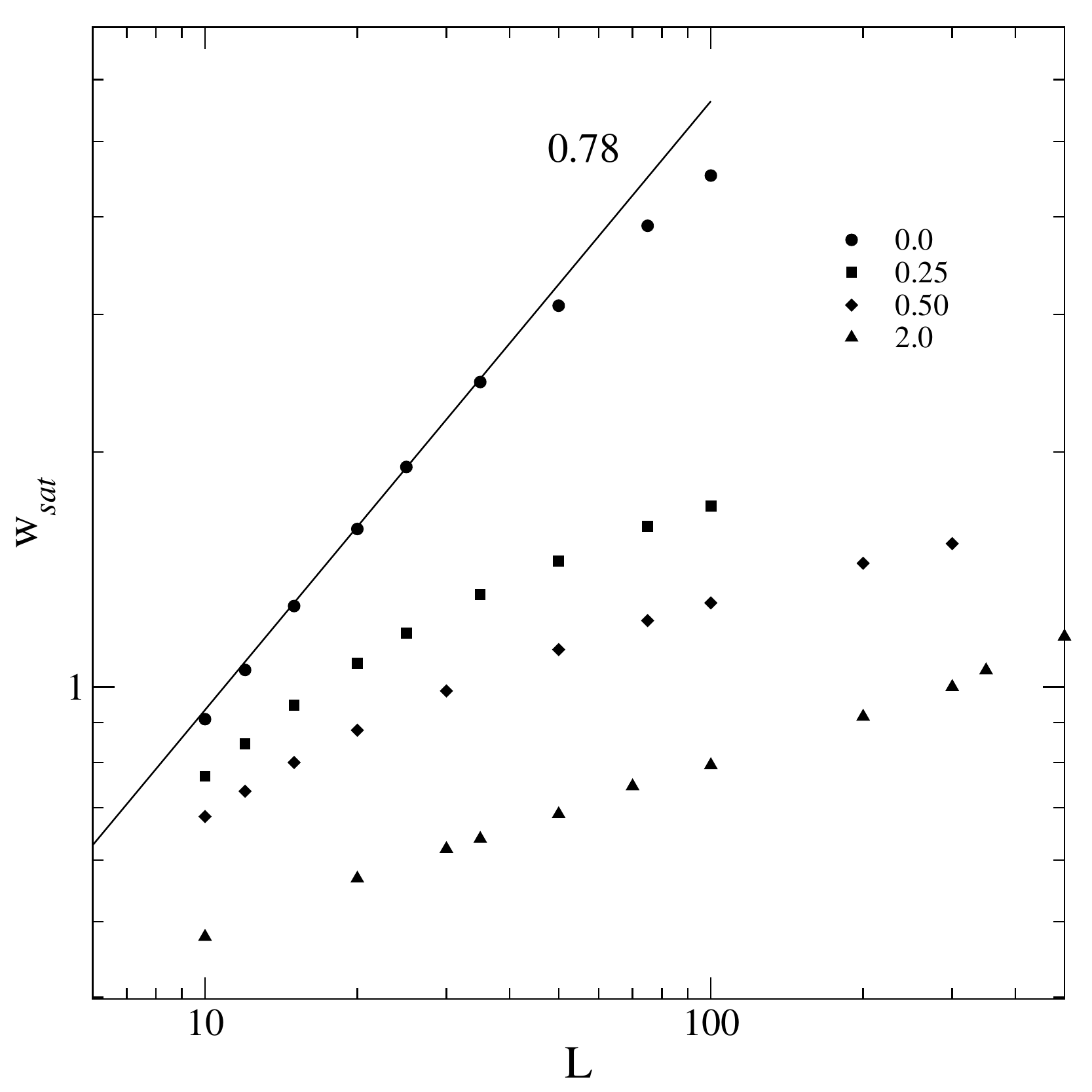}}~~~~~~~
\subfigure[\label{wsat2d_nuslog}]{\epsfxsize=6.5cm\epsfbox{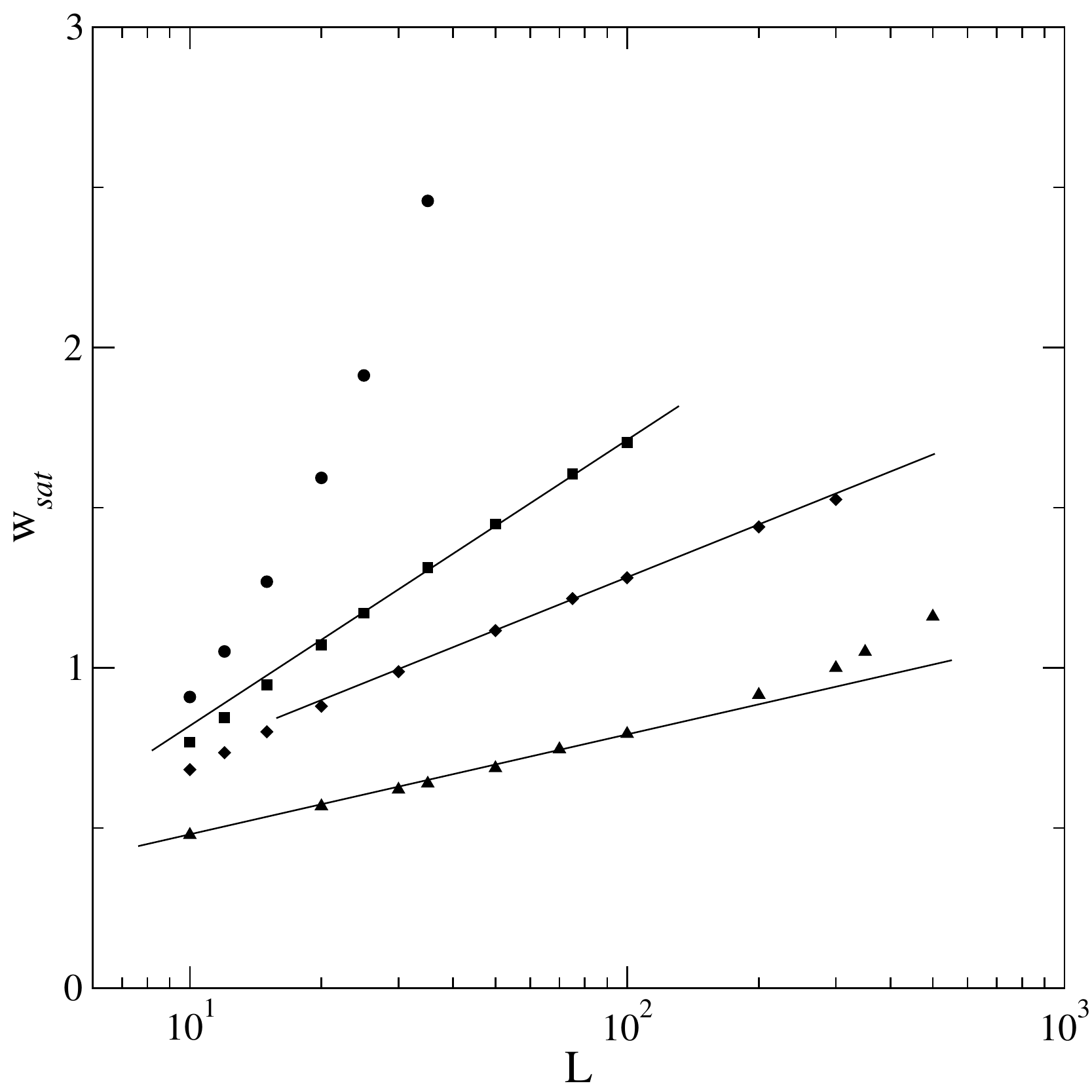}}
\end{center}
\caption{\label{wsat2d}\footnotesize Saturation roughness $w_{sat}$ as a function of the system size $L$ for distinct $\nu$ values in a $(2+1)$-dimensions system. In (a), a log-log plot is shown and, in (b), a semi-log plot. }
\end{figure}

In $(2+1)$-dimensions we obtained a similar scene as that observed in $(1+1)$-dimensions. The figures~\ref{rugo2d} and \ref{wsat2d} show the log-log and the semi log plots for the time evolution of the roughness and the system size dependency of the saturation roughness, respectively.
As can be seen in these figures, for values of $\nu\le0.5$ a crossover from a power law to a logarithm growth was observed for both, the roughness and the saturation roughness scaling. The initial value obtained to the growth ($\beta\approx 0.22$) and roughness ($\alpha\approx 0.78$) exponents are very close to those found to the universality class of the $\nabla^4h$ term, for wich $\beta = 1/5$ and $\alpha = 2/3$ and the asymptotic behavior is logarithmic (the growth and roughness exponent in the EW universality class are null that indicates a logarithmic behavior). For larger values of $\nu$, the scaling is always logarithmic.

The results observed using intermediate $\nu$ values for both, $1+1$ and $2+1$ dimensions, indicating the EW asymptotic universality class are in good agreement with results found on literature~\cite{Krug, Krug1, Vvedensky, Vvedensky1, Smilauer}. Besides, the time to the asymptotic behavior appear decreases as $\nu$ is increased, so intermediate values of $\nu$ can be used to anticipate the asymptotic universality class of the WV model.

\section{Conclusions}

In this work we studied a modified Wolf-Villain growth model with a probabilistic growth rule proportional to the power of the coordination number of a growing site, {\it i.e.}, $p \propto n^\nu$. Simulations in $(1+1)$-dimensions indicate that, for low $\nu$ values, the model presents a crossover from the $\nabla^4h$ to the $\nabla^2h$ behavior, in agreement with the original model. We observe that the crossover time separating these two behavior decreases as $\nu$ increases. This fact indicates that intermediary values of $\nu$ can be used to anticipate the asymptotic universality class of the WV model. However, for intermediate $\nu$ values, the model presents a crossover from the linear $\nabla^2h$ to the $(\nabla h)^2$ nonlinear behavior and, for larger $\nu$ values, the model falls into the KPZ universality class. The presence of the KPZ asymptotic universality class in this modified version can be easily explained. The growth probability in this version introduce a refusal of particle. As know in the literature \cite{KK}, the refusal of particle is one of the basic condition to the appearance of the KPZ universality class. In $(2+1)$-dimensions, the model presents the same type of crossover and, as in the $(1+1)$-dimensional case, the crossover time is dependent of the $\nu$ value. For larger value, we found that the model belongs to the EW universality class. In further work, we intend to apply the probabilistic rule to models, {e.g.} the DT model, that take into account the relaxation to the neighbor considering its coordination number.

\textbf{Acknowledgments}\\~\\ We thank to  A. P. F. Atman by the critical reading of the manuscript. This work was supported by CNPq, CAPES, and FAPEMIG Brazilian agencies.


\begin{thebibliography}{99}

\bibitem{Barabasi} Barab\'asi  A-L and Stanley H E, 1995 {\it Fractal concepts in surface growth}, (Cambridge University Press, Cambridge).

\bibitem{Meakin} Meakin P, {\it Fractals, scaling and growth far from equilibrium},
(Cambridge University Press, Cambridge, 1998).

\bibitem{fv} Family F and Vicsek T, 1985 J. Phys. A: Mathematical and General {\bf 18}, L75

\bibitem{Lai} Lai Z-W and Das Sarma S, 1991 Phys. Rev. Lett. {\bf 66}, 2348

\bibitem{kpz} Kardar  M, Parisi G and Zhang  Y C, 1986 Phys. Rev. Lett. {\bf 56}, 889

\bibitem{ew} Edwards S F and Wilkinson D R, 1982 Proc. R. Soc. Lond. \textbf{A381}, 17

\bibitem{wv} Wolf D E and Villain J, 1990 Europhys. Lett. \textbf{13}, 389

\bibitem{sarma-tamborenea} Das Sarma S and Tamborenea P, 1991 Phys. Rev. Lett.  {\bf 66} 325

\bibitem{tales} da Silva  T J and Moreira J G, 2002 Phys. Rev. E  {\bf 66} 061604

\bibitem{Krug} Krug J, Plischke M, and Siegert M, 1993 Phys. Rev. Lett. \textbf{70}, 3271

\bibitem{Krug1} Krug J, Plischke M and Siegert M, 1993 Phys. Rev. Lett. \textbf{71}, 949

\bibitem{Das_NR} Das Sarma S, Chatraphron p and Toroczkai Z, 2002 Phys. Rev. E {\bf 65} 036144

\bibitem{Costa} Costa B S, Euz\'ebio  J A R and Aar\~ao Reis F D A, 2003 Phys. A {\bf 328} 193

\bibitem{Vvedensky} Vvedensky D D, 2003 Phys. Rev. E {\bf 68}, 010601

\bibitem{Vvedensky1} Haselwandter C A and Vvedensky D D, 2007 Phys. Rev. E {\bf 76}, 041115

\bibitem{Paiva} Paiva L R and Ferreira S, 2007 J. Phys. A: Math. Theor. {\bf 40} 43

\bibitem{Bogoyavlenskiy} Bogoyavlenskiy V A, 2002 J. Phys. A: Math. Gen. {\bf 35} 2533

\bibitem{Alves_bogo} Alves S G and Ferreira S C, 2006 J. Phys. A: Math. Gen. {\bf 39} 2843

\bibitem{KK} Kim J M and Kosterlitz J M, 1989 Phys. Rev. Lett. \textbf{62}, 2289

\bibitem{Ryu} Ryu C S and Kim I M, 1995 Phys. Rev. E \textbf{51}, 3069

\bibitem{Smilauer}$\check{S}$milauer P and Kotrla M, 1994 Phys. Rev. B \textbf{49}, 5769

\end{thebibliography}
\end{document}